\documentclass[runningheads]{llncs}
\usepackage[T1]{fontenc}
\usepackage{graphicx}
\usepackage{hyperref}
\usepackage{amsmath,amsfonts}
\usepackage{algorithm}
\usepackage{algorithmic}
\usepackage{textcomp}
\usepackage[dvipsnames]{xcolor}
\usepackage{multirow}
\usepackage{caption}
\usepackage{subcaption}
\usepackage{float}
\usepackage{booktabs}
\usepackage{makecell}

\usepackage{pifont}

\newcommand{\greencheck}{{\color{ForestGreen}\ding{51}}}%
\newcommand{\xmark}{{\color{red}\ding{55}}}%

\def\BibTeX{{\rm B\kern-.05em{\sc i\kern-.025em b}\kern-.08em
    T\kern-.1667em\lower.7ex\hbox{E}\kern-.125emX}}
\begin{document}
\title{Cluster-Aware Attacks on Graph Watermarks}
%
%
\author{Alexander Nemecek\inst{1}\orcidID{0009-0000-2179-4186} \and
Emre Yilmaz\inst{2}\orcidID{0000-0003-0834-8805} \and
Erman Ayday\inst{1}\orcidID{0000-0003-3383-1081}}
\authorrunning{A. Nemecek et al.}
%
\institute{Case Western Reserve University, Cleveland, OH 44106, USA \\
\email{\{ajn98,exa208\}@case.edu}
\and
University of Houston-Downtown, Houston, TX 77002, USA\\
\email{yilmaze@uhd.edu}\\
}
\maketitle              
\begin{abstract}
Graph-structured datasets are increasingly central to sensitive applications spanning social networks, biomedical research, and cryptographic systems. As organizations share these datasets with trusted parties for collaborative analysis, protecting against unauthorized redistribution becomes critical. Graph watermarking addresses this challenge by embedding detectable signatures that enable ownership verification and attribution of leaked data. However, despite advances in watermarking techniques, existing robustness evaluations remain limited to random edge perturbation attacks, overlooking more sophisticated adversaries who exploit community structure present in real-world graphs. We introduce the first systematic evaluation of cluster-aware attacks on graph watermarking schemes. We present a threat model in which adversaries leverage community detection algorithms to guide strategic edge modifications, targeting either intra-cluster densification with inter-cluster boundary removal, or intra-cluster sparsification with inter-cluster noise injection. Evaluating against representative structural and spectral watermarking schemes, we demonstrate that cluster-aware attacks outperform random perturbations across real-world datasets and clustering algorithms. Our findings reveal that cluster-aware attacks reduce attribution accuracy while introducing structural distortion comparable to random attacks in most configurations, demonstrating superior attack efficiency. These results establish that current watermarking schemes, evaluated solely against random perturbations, remain vulnerable to structure-aware adversarial behavior, highlighting the need for robust defenses that account for community-exploiting adversaries in graph-based systems. Our code is available at \url{https://github.com/ANCP2021/Cluster-Aware-Graph-Watermarking-Attacks}.

\end{abstract}
\section{Introduction}\label{introduction}
Many datasets are structured as graphs, representing relationships in many domains, including social networks~\cite{scott2011social}, biomedical research~\cite{li2022graph}, and cryptographic applications~\cite{amudha2018application}. These graphs often contain sensitive information, making their secure storage, sharing, and analysis a concern. Researchers and organizations frequently share graph-structured data with trusted parties for collaborative analysis and real-world applications. However, the sensitive nature of this data raises security and privacy risks, including unauthorized access and data leakage among distributed entities. To mitigate these threats, data owners must ensure their datasets remain protected against adversarial modifications and that any unauthorized redistribution can be attributed to its source.

Watermarking, the process of embedding a detectable signature within an object, has been widely studied across multiple domains, including imaging~\cite{nikolaidis1998robust}, audio~\cite{bassia2001robust}, databases~\cite{ji2023privacy}, and machine learning models~\cite{nemecek2024topic,you2024gnnguard}. 
Graph-structured data is no exception; in a graph watermarking, signatures are embedded through structural modifications~\cite{zhao2015towards,eppstein2016models} or spectral transformations~\cite{bourree2025fast}, enabling ownership verification and unauthorized redistribution detection. However, despite advances in embedding techniques, the robustness evaluation of these schemes remains limited. Specifically, existing work primarily evaluates robustness against random edge perturbation attacks, failing to consider more sophisticated, structure-aware adversaries.

Real-world graphs typically exhibit community structure, where nodes naturally form densely connected subgroups (clusters). An adversary who recognizes these structures can develop targeted attack strategies that selectively modify edges within or between clusters to compromise graph watermark integrity more effectively than random edge modifications. This structural vulnerability is independent of the watermarking technique employed: both structural and spectral approaches rely on preserving statistical properties that community-aware perturbations can disrupt. Such adversaries are not hypothetical, as an insider at a recipient organization with only simple network science knowledge can run community detection on a leaked graph using freely available libraries (e.g., NetworkX or igraph), with no access to the watermarking system or feedback from it. Analogous structure-aware strategies are already documented in adjacent literature, where community-guided perturbations degrade graph neural networks more effectively than random baselines~\cite{dai2018adversarial}. Despite this low barrier to entry and documented precedent, no existing work evaluates watermarking robustness against cluster-aware adversaries. All prior schemes are evaluated exclusively against random edge perturbations, leaving data owners with a false sense of attribution guarantees against a realistic class of adversaries.

The goal of this work is to address the gap in existing literature by introducing the first systematic evaluation of cluster-aware attacks on graph watermarking. We present a threat model in which adversaries leverage community detection algorithms to guide strategic edge modifications, and we demonstrate that such attacks are more effective than random baselines across multiple graph datasets and clustering algorithms. Our key contributions can be summarized as follows:
\begin{enumerate}
    \item \textbf{Cluster-Aware Threat Model}: We introduce a novel adversarial 
    model in which attackers exploit graph community structure through two 
    strategic approaches: intra-cluster densification with inter-cluster 
    boundary removal, and intra-cluster sparsification with inter-cluster 
    noise injection.
    
    \item \textbf{Evaluation Across Watermarking Paradigms}: We evaluate cluster-aware attacks against both structural~\cite{zhao2015towards} and spectral~\cite{bourree2025fast} watermarking schemes, demonstrating that vulnerability to structure-aware adversaries generalizes across embedding strategies.
    
    \item \textbf{Parameter-Free Clustering Analysis}: We evaluate four 
    parameter-free community detection algorithms, providing the first systematic analysis of clustering method effectiveness for adversarial graph perturbation under realistic black-box conditions.
    
    \item \textbf{Empirical Findings}: Across three real-world graphs with varying structural properties and both watermarking paradigms, we show that cluster-aware attacks reduce attribution accuracy compared to random edge flipping while introducing comparable structural distortion across most configurations, demonstrating superior attack efficiency.
\end{enumerate}

As graph-based data becomes increasingly critical to both the academic community and industry, ensuring the ability to trace and protect such data against evolving attack strategies is paramount. Our findings suggest that current watermarking schemes, evaluated solely against random perturbations, remain vulnerable to structure-aware adversarial behavior. By addressing these gaps, we aim to establish a more robust benchmark for future graph watermarking designs.

\section{Background and Related Work}\label{related}
In this section, we review the background on graph watermarking schemes, community detection in graphs, and existing threat models evaluated in prior work.

\subsection{Graph Watermarking}\label{graph-watermarkin}
Graph watermarking is a subfield of digital watermarking that focuses on embedding detectable signatures within graph-structured data to enable ownership verification and unauthorized redistribution detection. The process transforms an original graph $G$ into a watermarked version $G'$ by embedding a user-specific signature that can be extracted and verified from a potentially modified graph $\hat{G}$, even after adversarial perturbations. 

Despite the increasing importance of graph-structured data, research on graph watermarking remains relatively limited. To the best of our knowledge, only three watermarking schemes have been proposed in the literature: Zhao et al.~\cite{zhao2015towards}, Eppstein et al.~\cite{eppstein2016models}, and Bourr\'ee et al.~\cite{bourree2025fast}. These schemes can be broadly categorized into two classes based on their embedding approach: \emph{structural}, which modify graph topology directly, and \emph{spectral}, which embed watermarks in the frequency domain of the graph's adjacency matrix.

\textbf{Structural watermarking} schemes embed signatures by directly modifying the topology of the graph through edge modifications. These methods typically select a subset of nodes and edges in the original graph $G$ and transform their local structure to encode a watermark pattern. The key principle is to create detectable topological features that can be regenerated and verified using a secret key, while minimizing distortion to the graph's global properties.

The most widely recognized structural graph watermarking scheme, proposed by Zhao et al., embeds a randomly generated Erd\H{o}s-R\'enyi subgraph watermark $W$ into the original graph $G$ by selecting $k$ nodes based on Node Structure Descriptors (NSDs)\footnote{A Node Structure Descriptor is a structural signature of a node defined as a sorted array of its neighbors' degrees. For example, a node with three neighbors of degrees 2, 6, and 4 has NSD label ``2-4-6.''} and performing XOR-based edge flipping to encode the watermark pattern. Extraction regenerates $W$ using a secret key, identifies candidate nodes in a potentially perturbed graph $\hat{G}$ by matching NSD labels, and performs subgraph matching to verify the watermark's presence. Successful extraction therefore requires both that watermark-carrying nodes retain NSD labels consistent with $W$'s expected structure, and that the edge pattern connecting them survives intact.

Building on Zhao et al., Eppstein et al. proposed a theoretical framework formalizing security guarantees for structural graph watermarking under Erd\H{o}s-R\'enyi and power-law random graph models, providing provable bounds on adversarial advantage.

\textbf{Spectral watermarking} schemes operate in the frequency domain rather than directly modifying graph topology, treating the graph's adjacency matrix as a signal and applying transformations to embed watermarks in specific frequency components. The Fast\&Fourier (F\&F) scheme proposed by Bourr\'ee et al. adapts image watermarking techniques to the graph domain by treating the adjacency matrix of $G$ as a two-dimensional signal. F\&F embeds a watermark key sampled from a Gaussian distribution into the lowest magnitude coefficients of the Fourier-transformed adjacency matrix, then applies the inverse transform and binarization to produce the watermarked graph. Extraction computes the spectral difference between a suspect graph $\hat{G}$ and the original graph $G$, verifying the watermark via a distance-based similarity threshold. The Fourier spectrum of the adjacency matrix encodes the graph's global structural patterns, which in graphs with community structure correspond to inter- and intra-community edge densities, the same properties targeted by cluster-aware perturbations. Unlike Zhao et al.'s strict subgraph-matching extraction, F\&F extraction is natively threshold-based, declaring a suspect graph watermarked when spectral similarity to the known key exceeds a configured threshold. F\&F offers computational advantages over structural methods, achieving $O(N^2 \log N)$ complexity compared to NP-complete subgraph matching operations.

\subsection{Community Detection in Graphs}
Clustering refers to the process of partitioning a set of objects into groups based on a similarity metric. In the context of graphs, clustering or community detection aims to identify densely connected subgroups of vertices (nodes) that are sparsely connected to the rest of the graph. Given a graph $G=(V, E)$ where $V$ is the set of vertices and $E$ is the set of edges, clustering divides $V$ into disjoint subsets $\mathcal{C}=\{C_1, C_2, \dots, C_k\}$ such that vertices within each cluster $C_i$ are more densely connected to one another than to vertices outside the cluster. Community structures are an inherent property of many real-world graphs and algorithms for detecting communities in graphs can be broadly categorized based on whether they require explicit hyperparameter tuning.

\emph{Parameter-free} methods operate without user-specified configuration, autonomously optimizing an objective function or converging to a stable partitioning~\cite{newman2006modularity}. These methods are particularly attractive for adversarial settings where attackers lack access to ground-truth information or feedback from the watermarking system. \emph{Parameterized} methods offer greater flexibility and control but require users to specify structural assumptions or resolution parameters~\cite{ng2001spectral}. While these methods can achieve superior partitioning quality when properly configured, their effectiveness depends on appropriate parameter selection, which may require domain expertise, iterative optimization, or access to reference graphs that are unlikely to be available to a black-box adversary. We therefore focus on parameter-free methods in our evaluation.

Beyond analytical use cases, community detection has been leveraged in adversarial contexts to exploit structural vulnerabilities in graph-based systems. In deanonymization attacks, adversaries analyze community structures and cross-reference them with auxiliary datasets to re-identify anonymized users~\cite{narayanan2006break}. In adversarial machine learning, clustering has guided targeted attacks on graph neural networks, where perturbations are strategically applied within or across communities to degrade model performance while minimizing detectable distortion~\cite{dai2018adversarial}. These examples demonstrate that community structure, while useful for analysis, can be exploited by adversaries who recognize that topology-aware perturbations are more effective than random modifications. This adversarial precedent motivates our investigation of whether community detection similarly benefits attackers against graph watermarking schemes.

\subsection{Threat Models in Graph Watermarking}
The security of graph watermarking schemes is evaluated based on their robustness to adversarial modifications. Existing literature has primarily addressed three categories of threat models: random edge perturbations, collusion attacks, and graph deanonymization.

\textbf{Random Edge Perturbation.} The most common threat model assumes an adversary who randomly modifies the watermarked graph by flipping edges, defined as either adding non-existent edges or removing existing ones. This baseline attack is structurally naive where the adversary selects edge modifications uniformly at random without considering graph structure. All existing graph watermarking schemes~\cite{zhao2015towards,eppstein2016models,bourree2025fast} evaluate robustness against this threat model.

\textbf{Collusion Attacks.} In collusion-based threat models, multiple recipients collaborate to identify and remove watermarks by comparing their distinct watermarked versions of the same graph. By analyzing discrepancies across copies, colluding adversaries can infer the locations of embedded watermark structures. Zhao et al. is the only work to evaluate robustness against collusion, simulating scenarios where attackers apply majority voting to isolate watermark-specific modifications. However, such attacks require coordination among multiple parties and assume attackers can successfully deanonymize and align their graphs prior to comparison.

\textbf{Graph Deanonymization.} Zhao et al. also address deanonymization attacks within their collusion model, randomly relabeling node identifiers before distribution. However, they acknowledge this defense can be circumvented through structural deanonymization techniques. Adversaries can exploit high-degree nodes as structural anchors, matching them across graphs based on degree sequences and local connectivity patterns.

\begin{table}[h]
\centering
\scriptsize
\setlength{\tabcolsep}{5pt}
\begin{tabular}{lcccc}
\toprule
\textbf{Watermarking}
  & \makecell{\textbf{Random Edge}\\\textbf{Perturbation}}
  & \makecell{\textbf{Collusion}\\\textbf{Attacks}}
  & \textbf{Deanonymization}
  & \makecell{\textbf{Cluster}\\\textbf{Aware}} \\
\midrule
\textit{Zhao et al.}~\cite{zhao2015towards}            & \greencheck & \greencheck & \greencheck & \xmark \\
\textit{Eppstein et al.}~\cite{eppstein2016models}     & \greencheck & \xmark      & \xmark      & \xmark \\
\textit{Bourr\'ee et al.\ (F\&F)}~\cite{bourree2025fast} & \greencheck & \xmark      & \xmark      & \xmark \\
\bottomrule
\end{tabular}
\caption{Adversarial threat models addressed by existing graph watermarking schemes. Cluster-aware attacks remain unexplored across all prior work.}
\label{Lit}
\end{table}

\textbf{The Gap: Cluster-Aware Attacks.} 
While Zhao et al.'s work represents a more robust threat model as opposed to those considered by Eppstein et al. and Bourr\'ee et al., a critical gap remains in the graph watermarking literature. As summarized in Table~\ref{Lit}, all existing schemes evaluate robustness exclusively against adversaries who perform random edge perturbations, without exploiting underlying structural properties of the graph. This limitation is significant because real-world graphs exhibit pronounced clustering behavior, and adversaries aware of these patterns may craft more effective attacks than random perturbations alone. We address this gap by introducing the first cluster-aware threat model for graph watermarking. We focus on the single-attacker scenario, as random perturbation serves as the common evaluation baseline across all existing schemes, enabling direct comparison of attack effectiveness. Our objective is to establish that cluster-aware strategies are more effective than random perturbations, exposing vulnerabilities in current watermarking designs and motivating the need for structure-aware defenses.

\section{Cluster-Aware Threat Model}\label{threat}
Prior evaluations of graph watermarking robustness, including those in Zhao et al.~\cite{zhao2015towards}, primarily focus on random edge perturbation or collusion-based attacks. These models assume adversaries act with limited insight into the graph's topology. In contrast, we introduce a threat model in which an attacker leverages the graph's inherent community structure to launch more targeted attacks. While structure-aware adversaries are established in adjacent graph problems such as deanonymization~\cite{narayanan2006break} and graph neural network attacks~\cite{dai2018adversarial}, no existing graph watermarking threat model exploits such structural properties (e.g., community structure, degree distributions) to guide edge modifications. To the best of our knowledge, this is the first study to model a cluster-aware adversary who leverages community detection to selectively add or remove edges in a way that degrades watermark integrity while preserving overall graph utility.

\subsection{Deployment Scenario}\label{deployment-scenario}
Consider a platform operator that shares a social interaction graph with several data analytics firms for collaborative research, embedding a distinct watermark in each firm's copy so that any leak can be traced to the responsible recipient. An employee at one such firm, motivated by competitive advantage, decides to leak their copy to an external party. Before redistributing it, the employee runs a community detection algorithm such as Louvain or Leiden using freely available libraries like NetworkX or igraph, requiring only a few lines of Python and simple network science knowledge. Armed with the resulting partitioning, the employee performs targeted edge modifications within and across communities to degrade the embedded watermark, then leaks the perturbed graph. When the data owner discovers the leak and attempts watermark extraction, the attempt fails: the owner cannot determine which firm was responsible, and the attribution guarantee that motivated watermarking collapses.

This scenario aligns with documented patterns of misuse. Insiders at partner organizations have exfiltrated large-scale relational datasets containing rich user-interaction data, including reported incidents in which third-party marketing or analytics firms leaked hundreds of millions of records from major platforms~\cite{schneble2018cambridge}. Such datasets are routinely analyzed as graphs in downstream use, making community detection a standard analytic step rather than a specialized research capability. In the adjacent domain of graph neural networks, Dai et al.~\cite{dai2018adversarial} demonstrated that structure-guided edge perturbations degrade model performance substantially more than random modifications, confirming that structure-aware adversarial behavior against graph-based systems is an established attack paradigm rather than a theoretical concern.

\subsection{Adversarial Assumptions}
We consider a standard watermarking scenario in which a data owner embeds user-specific watermarks before distributing individualized copies to multiple recipients. One recipient, the \textit{leaker}, maliciously redistributes their copy; the remaining recipients are \textit{non-leakers}. We assume a single-attacker model without collusion, isolating the effectiveness of cluster-aware perturbations as a standalone attack vector.

The adversary operates in a black-box setting with full access to their watermarked graph $G'$, but does not possess the original graph $G$, the watermark generation key, or the user-specific signature key, and cannot query the detection or extraction system. They therefore cannot regenerate the embedded watermark, identify the subgraph in which it was placed, or receive feedback on whether their perturbations evade detection. The attacker must act preemptively based solely on structural properties observable in $G'$, exploiting community modularity and local density patterns computable from $G'$ itself under the assumption that meaningful community structure survives watermark embedding. Their primary goal is to prevent successful extraction and attribution; a secondary objective is to preserve graph utility, introducing minimal distortion to the overall structure.

\section{Proposed Cluster-Aware Attack Strategy}\label{proposed}
We describe our proposed attack strategy in which an adversary exploits inherent community structure present in real-world graphs. The attacker begins by applying a community detection algorithm to the watermarked graph $G'$ to identify densely connected subgroups. Once communities are identified, the attacker selectively adds or removes edges based on their structural role on whether the edge lies within a community (intra-cluster) or spans different communities (inter-cluster). We define two attack strategies that exploit both dimensions:
\begin{enumerate}
\item \textbf{Strategy I (Intra-Add/Inter-Remove)}: Densifies communities via intra-cluster edge addition while weakening inter-community boundaries through edge removal.
\item \textbf{Strategy II (Intra-Remove/Inter-Add)}: Sparsifies communities through intra-cluster edge removal while injecting noise across community boundaries through inter-cluster edge addition.
\end{enumerate}

Let the clustering map $C$ assign each node of $G'$ to its community, so that $C^{-1}(c)$ is the set of nodes in community $c$. The attacker then iteratively performs two strategy-dependent operations with equal probability until a predefined perturbation budget is exhausted. For Strategy I, these are \textbf{intra-cluster addition} (a cluster is selected at random and an edge is added between two unconnected nodes within it) and \textbf{inter-cluster removal} (an existing edge connecting nodes from different clusters is randomly selected and removed). For Strategy II, the operations are inverted: \textbf{intra-cluster removal} (an existing intra-cluster edge is randomly selected and removed) and \textbf{inter-cluster addition} (two unconnected nodes from different clusters are selected and an edge is added between them). Both procedures are formalized in Algorithm~\ref{algo1}.

The two strategies degrade watermarks through complementary mechanisms. Strategy I increases intra-cluster density, creating more locally similar neighborhoods that reduce the structural distinctiveness of watermark-carrying regions, while removing inter-cluster edges raises modularity but perturbs the boundary degree and density statistics on which the watermark signature depends, making it harder to distinguish from the surrounding topology. Strategy II inverts both operations: removing intra-cluster edges fragments watermark-carrying regions and degrades the local structural patterns or spectral coefficients used for detection, while adding inter-cluster edges injects structural noise across community boundaries and reduces modularity, making watermark-induced patterns harder to isolate from the perturbed graph.

\begin{algorithm}
\caption{Cluster-Aware Edge Flip Attack (Strategies I and II)}
\label{algo1}
\begin{algorithmic}[1]
\STATE \textbf{Input:} Graph $G' = (V, E)$, clustering map $\mathcal{C}$,
       number of flips $n$, strategy $s \in \{\text{I}, \text{II}\}$
\STATE \COMMENT{Strategy I: intra-add/inter-remove;\ \ Strategy II: intra-remove/inter-add}
\STATE $F \leftarrow 0$
\WHILE{$F < n$}
    \STATE Sample $\alpha \sim \mathcal{U}(0, 1)$;\ \ $\mathrm{intra} \leftarrow (\alpha < 0.5)$
    \IF{($\mathrm{intra}$ \AND $s = \text{I}$) \OR ($\neg\mathrm{intra}$ \AND $s = \text{II}$)}
        \IF{$\mathrm{intra}$}
            \STATE Select random cluster $c$;\ choose $u,v \in \mathcal{C}^{-1}(c)$ s.t. $(u,v) \notin E(G')$
        \ELSE
            \STATE Pick clusters $c \neq c'$;\ choose $u \in \mathcal{C}^{-1}(c),\ v \in \mathcal{C}^{-1}(c')$ s.t. $(u,v) \notin E(G')$
        \ENDIF
        \IF{such $u,v$ exist}
            \STATE Add edge $(u,v)$ to $G'$;\ $F \leftarrow F + 1$
        \ENDIF
    \ELSE
        \IF{$\mathrm{intra}$}
            \STATE Pick $(u,v) \in E(G')$ s.t. $\mathcal{C}(u) = \mathcal{C}(v)$
        \ELSE
            \STATE Pick $(u,v) \in E(G')$ s.t. $\mathcal{C}(u) \neq \mathcal{C}(v)$
        \ENDIF
        \IF{such $(u,v)$ exists}
            \STATE Remove edge $(u,v)$ from $G'$;\ $F \leftarrow F + 1$
        \ENDIF
    \ENDIF
\ENDWHILE
\STATE \textbf{return} $\hat{G}$
\end{algorithmic}
\end{algorithm}

\section{Experimental Evaluation}\label{experiment}
In this section, we describe the setup used to evaluate our cluster-aware attack strategies. We describe the evaluated watermarking schemes, introduce the graph datasets selected for this study, detail the clustering methods and attack configurations, and define the evaluation metrics used to assess attack effectiveness.

\subsection{Evaluated Watermarking Schemes}\label{params}
We evaluate our cluster-aware attack strategies against two representative graph watermarking schemes: the structural scheme of Zhao et al.~\cite{zhao2015towards} and the spectral Fast\&Fourier (F\&F) scheme of Bourr\'ee et al.~\cite{bourree2025fast}. Together, these cover both embedding paradigms identified in Section~\ref{related}.

For both schemes, we embed a single watermark per graph to provide a clear evaluation of attack effectiveness under a binary detection scenario where the watermark is either successfully extracted or not. This provides a more stringent test than multi-watermark schemes where partial extraction may still enable attribution, allowing us to isolate the impact of cluster-aware attacks compared to random perturbations. We adopt the watermark parameters specified by each scheme's original evaluation.

For Zhao et al., we adopt a strict extraction methodology requiring perfect recovery of the embedded watermark subgraph through exact subgraph matching. While Zhao et al. also propose more tolerant extraction strategies, strict matching eliminates false positives and provides a conservative evaluation framework for comparing attack effectiveness. For F\&F, we adopt the default spectral similarity threshold specified by Bourr\'ee et al.~\cite{bourree2025fast}, under which extraction succeeds when the distance between the suspect graph's spectral signature and the known watermark key falls below the configured threshold. Full parameter settings for both schemes, along with evaluation of Zhao et al.'s tolerant extraction configuration, are provided in Appendix~\ref{scheme-params}.

\subsection{Datasets}\label{datasets}
We evaluate our attacks on three real-world graph datasets from the SNAP network dataset collection~\cite{snapnets}, representing diverse network types with varying structural characteristics: \textit{Social Circles: Facebook}~\cite{leskovec2012learning}, a social network; \textit{CAIDA AS}~\cite{leskovec2005graphs}, an autonomous system graph; and \textit{ArXiv Astro Physics}~\cite{leskovec2007graph}, a scientific collaboration network. These datasets were selected to capture variation in size and average degree, structural factors that influence both the embedding behavior of the evaluated watermarking schemes and the effectiveness of community detection. For brevity, we refer to these datasets as \textit{Facebook}, \textit{CAIDA}, and \textit{ArXiv} throughout the paper.

\begin{table}[ht]
\centering
\scriptsize
\caption{Summary statistics for evaluated graph datasets.}
\begin{tabular}{lrrr}
\toprule
\textbf{Graph} & \textbf{Nodes} & \textbf{Edges} & \textbf{Average Degree} \\
\midrule
Facebook & 4,039  & 88,234  & 43.69 \\
CAIDA    & 26,475 & 53,381  & 4.03 \\
ArXiv    & 18,772 & 198,110 & 21.11 \\
\bottomrule
\end{tabular}
\label{tab:datasets}
\end{table}

\textit{Facebook} is a social graph derived from anonymized Facebook ego networks collected via survey participants. \textit{CAIDA} is an autonomous system graph derived from RouteViews BGP table snapshots. \textit{ArXiv} is a scientific collaboration network from the arXiv e-print repository covering papers submitted to the Astrophysics category. Summary statistics for all three datasets are reported in Table~\ref{tab:datasets}. We verified that all three datasets satisfy the node degree and subgraph density feasibility criteria specified by Zhao et al. for watermark embedding under our chosen parameters; F\&F imposes no equivalent feasibility constraint.

\subsection{Clustering Algorithms}\label{clustering}
The first step of our cluster-aware attack is community detection, in which the adversary partitions the watermarked graph into clusters. We evaluate four parameter-free algorithms, restricting to methods that require no hyperparameter tuning because a black-box attacker has no detection-system feedback to validate parameter choices: \textbf{Greedy Modularity}~\cite{newman2006modularity}, a hierarchical agglomerative algorithm that iteratively merges communities to maximize modularity; \textbf{Label Propagation}~\cite{raghavan2007near}, which iteratively updates each node's label to match its most frequent neighbor label until stabilization; \textbf{Leiden}~\cite{traag2019louvain}, which refines modularity-based partitions through local moves and aggregation to ensure internally well-connected communities~\cite{blondel2008fast}; and \textbf{Infomap}~\cite{rosvall2008maps}, an information-theoretic method that minimizes the description length of a random walk on the graph.

\subsection{Attack Configuration}\label{attack-config}
We evaluate the effectiveness of our cluster-aware attack strategies (see Section~\ref{proposed}) against a random edge flipping baseline, using the clustering algorithms described in Section~\ref{clustering}. In the random baseline, the adversary randomly selects pairs of nodes and applies an XOR operation: if an edge exists between the selected nodes, it is removed; otherwise, it is added.

Following Zhao et al.'s experimental protocol, we perform 10 independent runs for each edge flip level for both watermarking schemes to account for randomness in the attack process and watermark embedding. The perturbation budget is scheme-dependent, reflecting the different robustness regimes of each extraction mechanism.

For Zhao et al., we evaluate perturbation budgets within 0.1\% of total edges, with per-graph increments chosen to capture the transition from successful extraction to evasion. For \textit{Facebook}, due to its smaller size and high density, watermarks degrade rapidly, so we use finer-grained increments (1-9 flips). For \textit{CAIDA} and \textit{ArXiv}, we use coarser increments spanning up to 45 and 100 flips respectively. This range is sufficient because extraction success reaches 0\% for all attack methods well within this budget (Section~\ref{results}).

For F\&F, the default similarity threshold provides inherent tolerance to perturbation, requiring substantially larger budgets to evade detection. We evaluate perturbation budgets up to approximately 15\% of total edges for each graph to capture the full transition from successful extraction to evasion because F\&F's extraction curves are sharper than Zhao et al.'s, exhibiting a prolonged plateau followed by a rapid collapse. We sample a wide range of intermediate flip levels rather than concentrating increments at the low end, reporting all evaluated operating points in our results.

\subsection{Evaluation Metrics}\label{eval-metrics}
We evaluate our cluster-aware attack strategies across two dimensions: the success rate of watermark extraction under attack and the structural distortion introduced to the watermarked graph. These metrics collectively assess both the effectiveness of attacks in evading detection and their impact on graph utility and feasibility. As attacks are performed offline, we also provide runtime analysis of clustering and attack execution in Appendix~\ref{runtime-performance}.

\textbf{Extraction Success Rate.}
We measure extraction success rate as the metric for attack effectiveness, defined as the percentage of trials in which the embedded watermark is successfully recovered under each scheme's native extraction procedure: exact subgraph matching against the regenerated watermark for Zhao et al., and spectral similarity exceeding the configured threshold for F\&F. We note that F\&F's underlying extraction signal is itself a continuous spectral similarity score, thresholded per Bourr\'ee et al.'s specification; the binary outcome we report is the thresholded view of this continuous signal.

\textbf{Structural Distortion Metrics.} To assess the impact of attacks on graph structure, we evaluate three complementary distortion metrics that capture different aspects of structural change.

\textit{Edit Distance.} Edit distance quantifies the proportion of edge modifications between the watermarked graph $G'$ and attacked graph $\hat{G}$:
\begin{equation}
    \text{ED}(G', \hat{G}) = \frac{|E(G') \oplus E(\hat{G})|}{|E(G')|} \times 100\%
\end{equation}
where $\oplus$ denotes symmetric difference. This metric serves as a fairness check where random and cluster-aware attacks modifying the same number of edges should yield approximately equal edit distances, ensuring comparisons are made under equivalent perturbation budgets.

\textit{dK-2 Deviation.} 
The dK-2 series characterizes graph structure through the joint degree distribution where for each edge $(u,v)$, it records the degree pair $(\deg(u), \deg(v))$, producing a normalized distribution over all such pairs. This captures the degree sequence and degree correlations (e.g., whether high-degree nodes connect to other high-degree nodes). Following prior work~\cite{zhao2015towards,eppstein2016models}, we quantify structural distortion as the normalized Euclidean distance between the dK-2 distributions of the watermarked graph $G'$ and attacked graph $\hat{G}$:
\begin{equation}
    \text{dK-2}(G', \hat{G}) = \frac{\|V_{G'} - V_{\hat{G}}\|_2}{\|V_{G'}\|_2}
\end{equation}
where $V_{G'}$ and $V_{\hat{G}}$ are the vectorized dK-2 probability distributions. Unlike edit distance, dK-2 deviation is sensitive to how attacks alter the statistical properties of the graph, not just the number of modifications.

\textit{Clustering Coefficient Change.}
The clustering coefficient change ($\Delta CC$) measures the relative change in global clustering coefficient between the watermarked graph $G'$ and attacked graph $\hat{G}$:
\begin{equation}
    \Delta CC = \frac{C(\hat{G}) - C(G')}{C(G')} \times 100\%
\end{equation}
where $C(G)$ is the global clustering coefficient (ratio of triangles to connected triples). This metric captures preservation of higher-order structure, specifically transitivity and triadic closure patterns that are critical for utility in social and collaboration networks.

\begin{figure*}[t!]
\centering
\includegraphics[width=0.75\textwidth]{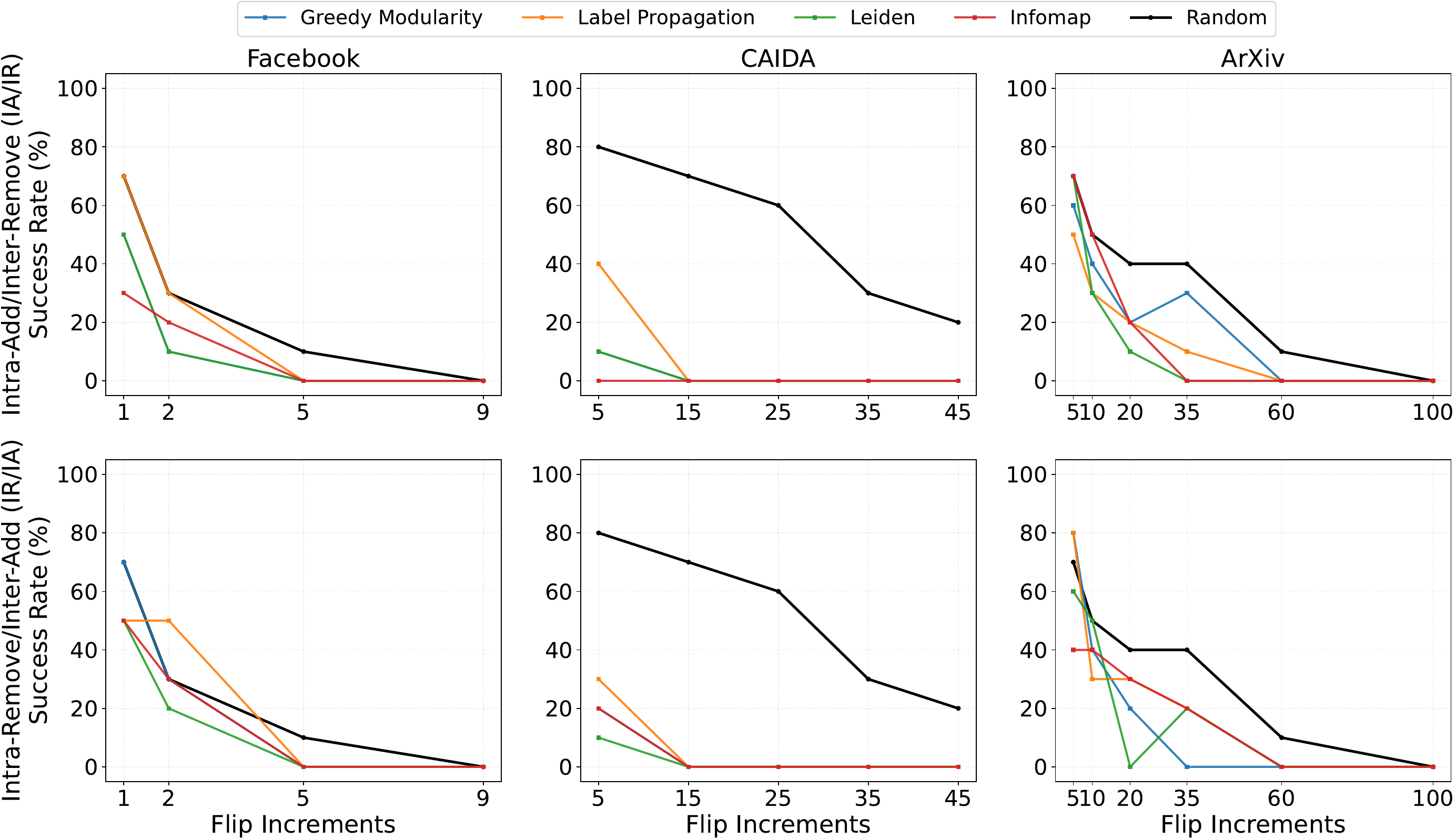}
\caption{Extraction success rate against Zhao et al. across datasets and attack strategies.}
\label{fig:zhao_success}
\end{figure*}

\begin{figure*}[t!]
\centering
\includegraphics[width=0.75\textwidth]{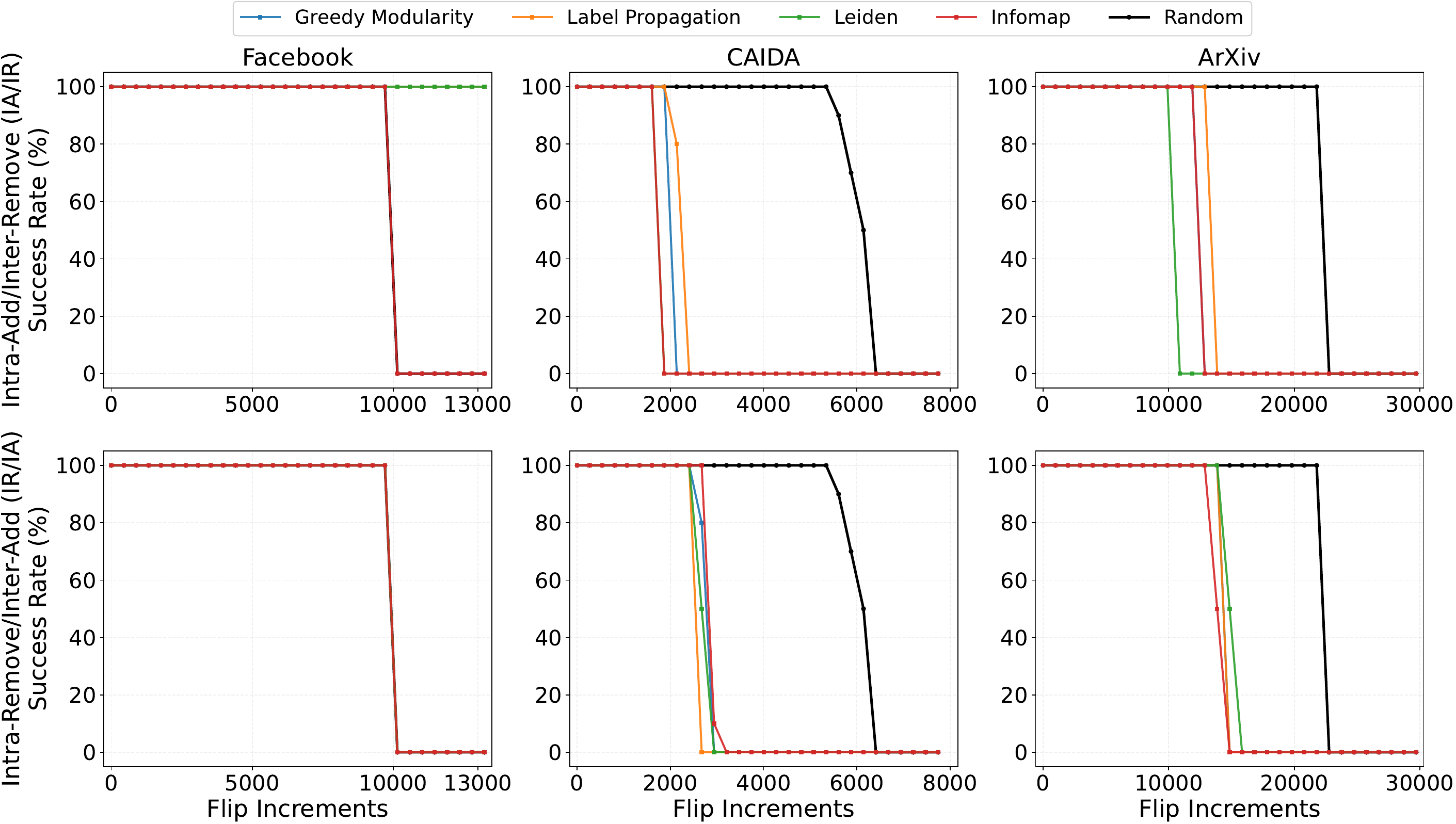}
\caption{Extraction success rate against F\&F across datasets and attack strategies.}
\label{fig:ff_success}
\end{figure*}

\section{Results}\label{results}
We present results comparing our two cluster-aware attack strategies (Intra-Add/Inter-Remove and Intra-Remove/Inter-Add) against the random edge flipping baseline across all datasets and clustering algorithms. We report extraction success rate and structural distortion metrics as defined in Section~\ref{eval-metrics}.

\subsection{Extraction Success Rate}\label{extraction-success}
Figures~\ref{fig:zhao_success} and~\ref{fig:ff_success} show watermark extraction success rates as a function of edge modifications for Zhao et al. and F\&F, respectively. Across both schemes, cluster-aware attacks consistently achieve lower extraction success with fewer edge modifications than the random baseline, demonstrating that exploiting community structure benefits adversaries regardless of the embedding paradigm.

Attack effectiveness varies with graph density (Table~\ref{tab:datasets}). On sparse \textit{CAIDA} (avg. degree 4.03), cluster-aware attacks show the clearest advantage under both schemes: against Zhao et al., they reduce extraction to 0-40\% at just 5 flip increments while random maintains $\sim$80\%; against F\&F, they collapse extraction near 2,500 flips versus $\sim$6,000 for random. \textit{ArXiv} (avg. degree 21.11) shows consistent but smaller advantages across both schemes, with cluster-aware methods reaching 0\% well before the random baseline. On dense \textit{Facebook} (avg. degree 43.69), high connectivity means even random flips disrupt structural signatures, reducing the relative advantage of cluster-aware methods against Zhao et al., though Greedy Modularity and Leiden still outperform random under Strategy I. For F\&F on \textit{Facebook}, most methods collapse near the $\sim$10\% perturbation mark, corresponding to F\&F's built-in spectral tolerance threshold (see Appendix~\ref{scheme-params}); Leiden is the only exception, where extraction is unchanged.

Among clusterings, Greedy Modularity and Leiden perform most consistently across datasets and schemes, reflecting their modularity-maximization objectives that identify structurally meaningful communities. Label Propagation fails to outperform random in only a few narrow operating points, all under Zhao et al. with the lowest flip counts on \textit{Facebook} (flip 1 for IA/IR and flip 2 for IR/IA) and flip 5 of \textit{ArXiv} IR/IA; elsewhere it tracks or beats random, and under F\&F it never underperforms random. We attribute this narrow underperformance to label oscillation producing unstable community assignments. When intra- and inter-cluster membership is noisy, Strategy I/II edge selections degenerate toward random flipping and lose the structural targeting that makes cluster-aware attacks effective (see Appendix~\ref{runtime-performance}).

\subsection{Structural Distortion}\label{structural-distortion-eval}
\begin{figure*}[h!]
\centering
\includegraphics[width=0.75\textwidth]{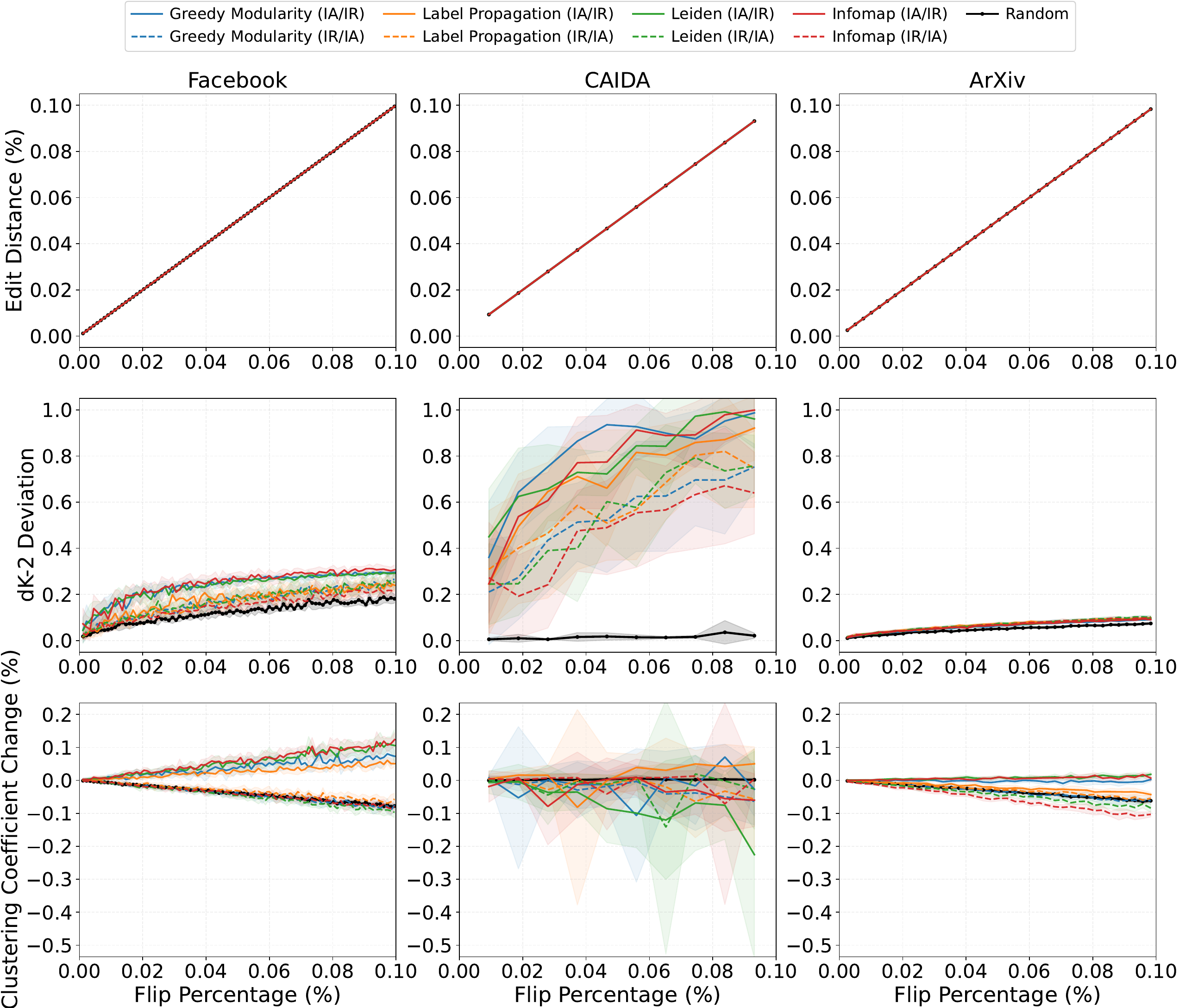}
\caption{Structural distortion metrics against Zhao et al. across datasets and attack strategies.}
\label{fig:zhao_structural_distortion}
\end{figure*}

\begin{figure*}[h!]
\centering
\includegraphics[width=0.75\textwidth]{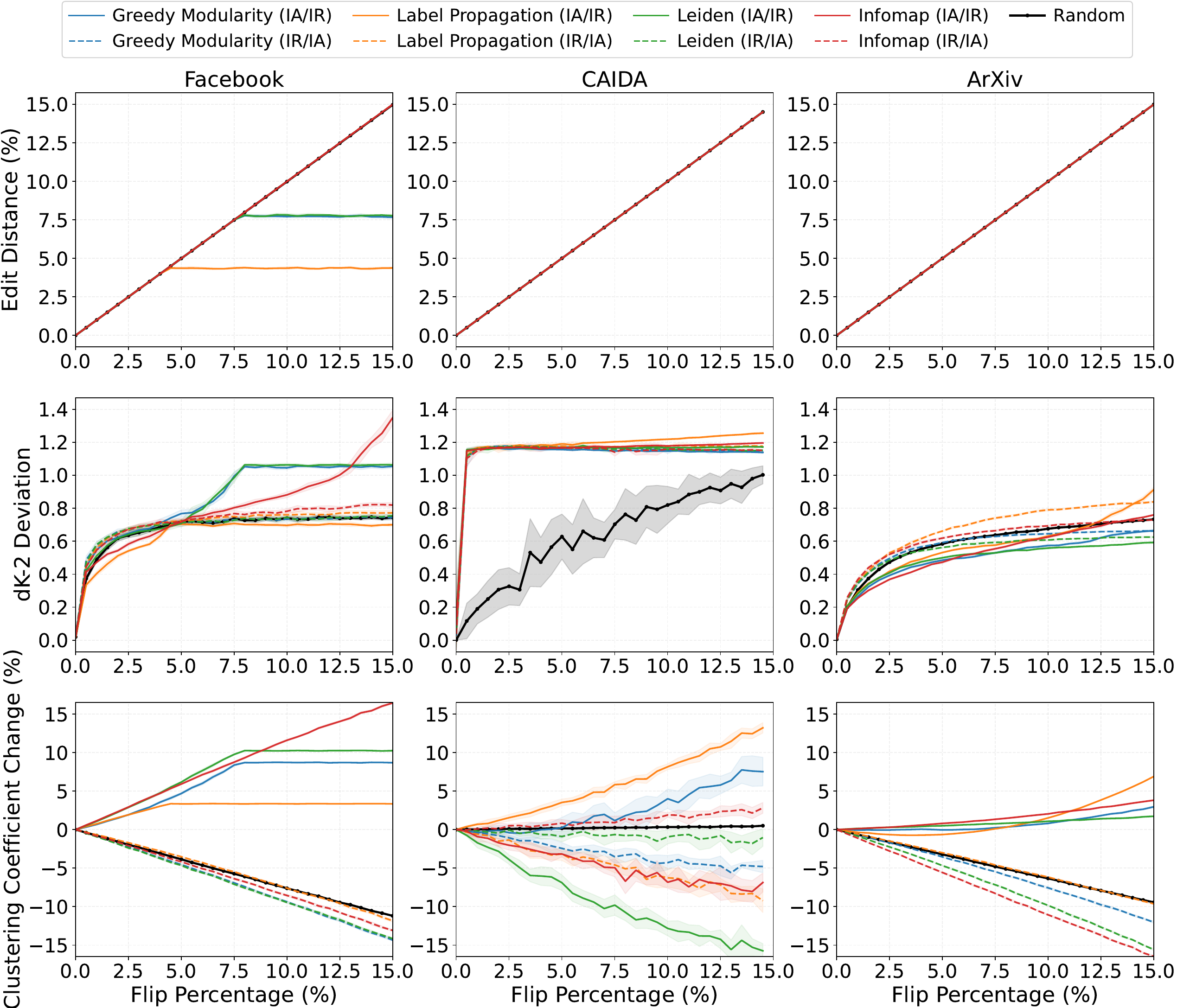}
\caption{Structural distortion metrics against F\&F across datasets and attack strategies.}
\label{fig:ff_structural_distortion}
\end{figure*}

To ensure fair comparison between attack strategies, we evaluate the structural impact of random and cluster-aware attacks on the watermarked graph. Figures~\ref{fig:zhao_structural_distortion} and~\ref{fig:ff_structural_distortion} show the distortion introduced against Zhao et al. and F\&F, respectively, measured as the difference between the watermarked graph $G'$ and the attacked graph $\hat{G}$.

\textbf{Edit Distance.} Edit distance scales linearly and overlaps across methods at matched budgets for all configurations, confirming equivalent perturbation budgets. The only exception is Strategy I (IA/IR) on \textit{Facebook} under F\&F, where Label Propagation, Greedy Modularity, and Leiden saturate near 5\% and 7.5\% respectively, because intra-cluster non-edges are exhausted; this occurs well after 0\% extraction success is reached and does not affect the fairness of comparisons in Section~\ref{extraction-success}.

\textbf{dK-2 Deviation.} We observe graph-dependent patterns that reveal a tradeoff between attack effectiveness and structural distortion, and this pattern holds across both watermarking schemes. For \textit{Facebook} and \textit{ArXiv}, dK-2 deviation for cluster-aware attacks is comparable to the random baseline at the perturbation budgets required to reach 0\% extraction success. \textit{CAIDA} presents a different pattern: cluster-aware strategies reach values near or above 1.0 while random baseline remains substantially lower, a gap that persists under both Zhao et al. and F\&F. This tradeoff reflects \textit{CAIDA}'s sparsity (average degree 4.03): random edge flips in sparse graphs are unlikely to alter the joint degree distribution because modifications are scattered, whereas cluster-aware attacks concentrate changes in structurally meaningful regions, disproportionately affecting degree correlations. Despite this increased distortion, cluster-aware attacks achieve near-complete watermark removal with fewer perturbations than random.

\textbf{Clustering Coefficient Change.} For \textit{Facebook} and \textit{ArXiv}, Strategy I increases the clustering coefficient (adding intra-cluster edges creates triangles) while Strategy II decreases it (removing intra-cluster edges breaks triangles); \textit{CAIDA} shows minimal deviation due to its low baseline clustering coefficient. Against Zhao et al., meaningful divergence from the random baseline occurs only after extraction success has already reached 0\%, indicating that cluster-aware attacks remove watermarks without dramatically altering triadic closure. Against F\&F, the larger perturbation budgets expose more pronounced divergence: Strategy I on \textit{ArXiv} and \textit{Facebook} produces noticeably higher clustering coefficient increases than random, and \textit{CAIDA} shows bidirectional divergence with Strategy I increasing and Strategy II decreasing the coefficient well beyond the random baseline. This reflects the extended operating range of F\&F attacks where cluster-aware strategies continue to reshape triadic structure beyond the point at which random attacks would have already degraded watermark integrity.

\section{Conclusion}
Graph-structured data is widely used across domains where maintaining attribution is critical. Graph watermarking enables this by embedding identifiable signatures into shared graph structures, yet remains understudied under cluster-aware threat models. We present the first evaluation of graph watermarks against cluster-aware adversaries who exploit the inherent community structure of real-world graphs. Evaluating against representative structural and spectral watermarking schemes, we demonstrate that cluster-aware attacks degrade detection performance more effectively than random perturbations while in most configurations introducing comparable structural distortion, with vulnerability to structure-aware adversaries generalizing across both embedding paradigms. Our findings reveal that current watermarking schemes, evaluated solely against random perturbations, fail to account for adversaries leveraging readily available community detection algorithms, emphasizing the urgent need for topological awareness in both watermark design and adversarial modeling. We discuss avenues for future work in Appendix~\ref{future-work}.

%
%
%
%
\bibliographystyle{splncs04}
\bibliography{references}

@article{scott2011social,
  title={Social network analysis: developments, advances, and prospects},
  author={Scott, John},
  journal={Social network analysis and mining},
  volume={1},
  pages={21--26},
  year={2011},
  publisher={Springer}
}

@article{li2022graph,
  title={Graph representation learning in biomedicine and healthcare},
  author={Li, Michelle M and Huang, Kexin and Zitnik, Marinka},
  journal={Nature Biomedical Engineering},
  volume={6},
  number={12},
  pages={1353--1369},
  year={2022},
  publisher={Nature Publishing Group UK London}
}

@article{amudha2018application,
  title={An application of graph theory in cryptography},
  author={Amudha, P and Sagayaraj, AC Charles and Sheela, AC Shantha},
  journal={International Journal of Pure and Applied Mathematics},
  volume={119},
  number={13},
  pages={375--383},
  year={2018}
}

@article{nikolaidis1998robust,
  title={Robust image watermarking in the spatial domain},
  author={Nikolaidis, Nikos and Pitas, Ioannis},
  journal={Signal processing},
  volume={66},
  number={3},
  pages={385--403},
  year={1998},
  publisher={Elsevier}
}

@article{bassia2001robust,
  title={Robust audio watermarking in the time domain},
  author={Bassia, Paraskevi and Pitas, Ioannis and Nikolaidis, Nikos},
  journal={IEEE Transactions on multimedia},
  volume={3},
  number={2},
  pages={232--241},
  year={2001},
  publisher={IEEE}
}

@inproceedings{ji2023privacy,
  title={Privacy-preserving database fingerprinting},
  author={Ji, Tianxi and Ayday, Erman and Yilmaz, Emre and Li, Ming and Li, Pan},
  booktitle={NDSS symposium},
  volume={2023},
  pages={10--14722},
  year={2023}
}

@article{nemecek2024topic,
  title={Topic-based watermarks for large language models},
  author={Nemecek, Alexander and Jiang, Yuzhou and Ayday, Erman},
  journal={arXiv preprint arXiv:2404.02138},
  year={2024}
}

@inproceedings{you2024gnnguard,
  title={Gnnguard: A fingerprinting framework for verifying ownerships of graph neural networks},
  author={You, Xiaoyu and Jiang, Youhe and Xu, Jianwei and Zhang, Mi and Yang, Min},
  booktitle={The Web Conference 2024},
  year={2024}
}

@inproceedings{zhao2015towards,
  title={Towards graph watermarks},
  author={Zhao, Xiaohan and Liu, Qingyun and Zheng, Haitao and Zhao, Ben Y},
  booktitle={Proceedings of the 2015 ACM on Conference on Online Social Networks},
  pages={101--112},
  year={2015}
}

@inproceedings{eppstein2016models,
  title={Models and algorithms for graph watermarking},
  author={Eppstein, David and Goodrich, Michael T and Lam, Jenny and Mamano, Nil and Mitzenmacher, Michael and Torres, Manuel},
  booktitle={Information Security: 19th International Conference, ISC 2016, Honolulu, HI, USA, September 3-6, 2016. Proceedings 19},
  pages={283--301},
  year={2016},
  organization={Springer}
}

@article{bourree2025fast,
  title={Fast In-Spectrum Graph Watermarks},
  author={Bourr{\'e}e, Jade Garcia and Kermarrec, Anne-Marie and Merrer, Erwan Le and Safsafi, Othmane},
  journal={arXiv preprint arXiv:2502.04182},
  year={2025}
}

@article{newman2006modularity,
  title={Modularity and community structure in networks},
  author={Newman, Mark EJ},
  journal={Proceedings of the national academy of sciences},
  volume={103},
  number={23},
  pages={8577--8582},
  year={2006},
  publisher={National Academy of Sciences}
}

@article{narayanan2006break,
  title={How to break anonymity of the netflix prize dataset},
  author={Narayanan, Arvind and Shmatikov, Vitaly},
  journal={arXiv preprint cs/0610105},
  year={2006}
}

@inproceedings{dai2018adversarial,
  title={Adversarial attack on graph structured data},
  author={Dai, Hanjun and Li, Hui and Tian, Tian and Huang, Xin and Wang, Lin and Zhu, Jun and Song, Le},
  booktitle={International conference on machine learning},
  pages={1115--1124},
  year={2018},
  organization={PMLR}
}

@article{ng2001spectral,
  title={On spectral clustering: Analysis and an algorithm},
  author={Ng, Andrew and Jordan, Michael and Weiss, Yair},
  journal={Advances in neural information processing systems},
  volume={14},
  year={2001}
}

@article{blondel2008fast,
  title={Fast unfolding of communities in large networks},
  author={Blondel, Vincent D and Guillaume, Jean-Loup and Lambiotte, Renaud and Lefebvre, Etienne},
  journal={Journal of statistical mechanics: theory and experiment},
  volume={2008},
  number={10},
  pages={P10008},
  year={2008},
  publisher={IOP Publishing}
}

@article{traag2019louvain,
  title={From Louvain to Leiden: guaranteeing well-connected communities},
  author={Traag, Vincent A and Waltman, Ludo and Van Eck, Nees Jan},
  journal={Scientific reports},
  volume={9},
  number={1},
  pages={1--12},
  year={2019},
  publisher={Nature Publishing Group}
}

@misc{snapnets,
  author       = {Jure Leskovec and Andrej Krevl},
  title        = {{SNAP Datasets}: {Stanford} Large Network Dataset Collection},
  howpublished = {\url{http://snap.stanford.edu/data}},
  month        = jun,
  year         = 2014
}

@article{leskovec2012learning,
  title={Learning to discover social circles in ego networks},
  author={Leskovec, Jure and Mcauley, Julian},
  journal={Advances in neural information processing systems},
  volume={25},
  year={2012}
}

@inproceedings{leskovec2005graphs,
  title={Graphs over time: densification laws, shrinking diameters and possible explanations},
  author={Leskovec, Jure and Kleinberg, Jon and Faloutsos, Christos},
  booktitle={Proceedings of the eleventh ACM SIGKDD international conference on Knowledge discovery in data mining},
  pages={177--187},
  year={2005}
}

@article{leskovec2007graph,
  title={Graph evolution: Densification and shrinking diameters},
  author={Leskovec, Jure and Kleinberg, Jon and Faloutsos, Christos},
  journal={ACM transactions on Knowledge Discovery from Data (TKDD)},
  volume={1},
  number={1},
  pages={2--es},
  year={2007},
  publisher={ACM New York, NY, USA}
}

@article{raghavan2007near,
  title={Near linear time algorithm to detect community structures in large-scale networks},
  author={Raghavan, Usha Nandini and Albert, R{\'e}ka and Kumara, Soundar},
  journal={Physical Review E—Statistical, Nonlinear, and Soft Matter Physics},
  volume={76},
  number={3},
  pages={036106},
  year={2007},
  publisher={APS}
}

@article{rosvall2008maps,
  title={Maps of random walks on complex networks reveal community structure},
  author={Rosvall, Martin and Bergstrom, Carl T},
  journal={Proceedings of the national academy of sciences},
  volume={105},
  number={4},
  pages={1118--1123},
  year={2008},
  publisher={National Academy of Sciences}
}

@article{schneble2018cambridge,
  title={The Cambridge Analytica affair and Internet-mediated research},
  author={Schneble, Christophe Olivier and Elger, Bernice Simone and Shaw, David},
  journal={The EMBO Reports},
  volume={19},
  number={8},
  pages={EMBR201846579},
  year={2018},
  publisher={Springer}
}

\appendix

\section{Future Work}\label{future-work}
\textbf{Alternative Structure-Preserving Attacks.} Cluster-aware perturbation is, to our knowledge, the first structure-aware attack evaluated against graph watermarking schemes. Additional structure-preserving strategies, such as targeting high-centrality nodes, exploiting motif-level statistics, or preserving global spectral properties while degrading embedded signatures, would broaden the adversarial landscape and provide tighter robustness benchmarks for future watermarking designs. 

\noindent\textbf{Stealthy Cluster-Aware Attacks.} An adversary concerned with evading structural monitoring (e.g., dK-2 drift or clustering coefficient change) could constrain edge selection to preserve such statistics through degree-preserving swaps within and across communities. Evaluating these stealthy variants of Strategies I and II would formalize the stealth-efficacy tradeoff, particularly for sparse graphs where the structural cost of targeted perturbation is most pronounced.

\section{Watermarking Scheme Parameters}\label{scheme-params}
This appendix specifies the parameter configurations used in our evaluation. We adopt the default settings proposed by each scheme's original authors~\cite{zhao2015towards,bourree2025fast} and additionally report results under Zhao et al.'s tolerant extraction variant.

\textbf{Zhao et al. Strict Configuration.}
We adopt the default watermark parameters recommended by Zhao et al.: edge probability $p=0.5$ and uniqueness constant $\delta=0.3$, yielding watermark size $k=(2+\delta)\log_2n$, where $n$ is the number of nodes in $G$. A single watermark is embedded per graph, and node identifiers are randomly relabeled post-embedding consistent with Zhao et al.'s procedure. The strict extraction configuration is enforced by setting the NSD degree bucket size $B=1$, the NSD label overlap threshold $\theta=1.0$, and the subgraph edit tolerance $L=0$, corresponding to exact NSD label matching and exact subgraph matching.

\textbf{F\&F Configuration.} We adopt the default F\&F configuration. The watermark key is sampled from a Gaussian distribution with standard deviation $\sigma$ and length $m$, embedded into the lowest-magnitude coefficients of the Fourier-transformed adjacency matrix, with extraction declaring a suspect graph watermarked when the spectral similarity distance falls below a threshold $\theta$. The key length is set as $m=\max(50, \lfloor 10 \cdot d \rceil)$, where $d=|E|/|V|$ is the graph density, mirroring the density-dependent key-length convention. The standard deviation $\sigma$ is selected via the dichotomous search procedure, which identifies the largest value producing a strictly positive edit distance while keeping distortion below the $10^{-2}$\% target.

The extraction threshold $\theta$ is calibrated per-graph to tolerate 10\% edge flips, matching the conservative upper bound. Specifically, we run 20 random-attack calibration trials at a perturbation level of $0.10 \cdot |E_{G'}|$ edge flips and set $\theta = 1.05 \cdot \max(s)$, where $s$ is the set of observed spectral similarity scores across those trials. This yields a threshold that tolerates random perturbation up to the calibration budget without producing false negatives on the watermarked graph. A single watermark is embedded per graph. Resolved parameter values for each dataset are produced at runtime from these rules.

\begin{figure*}[t]
    \centering
    \includegraphics[width=0.75\textwidth]{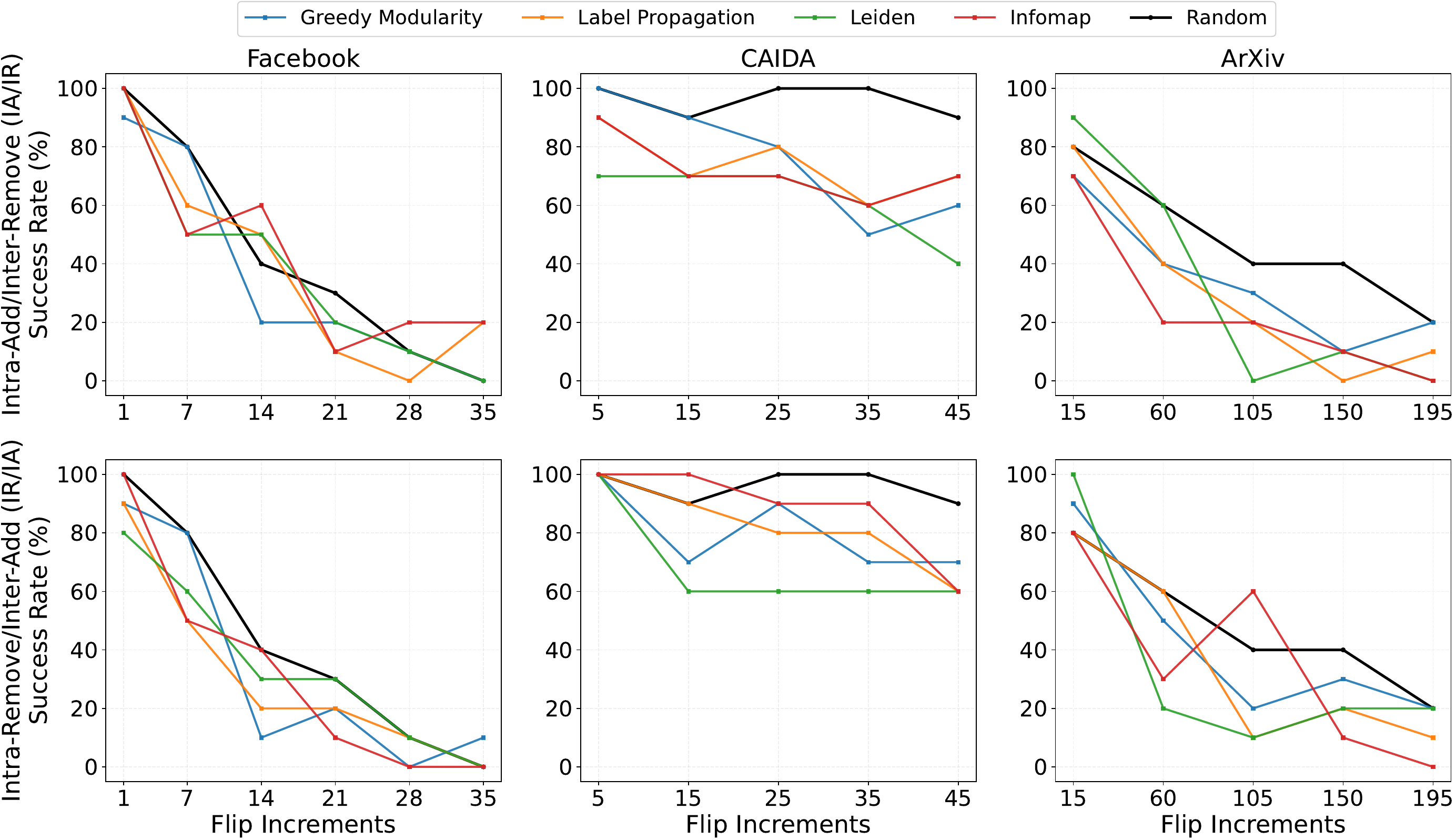}
    \caption{Extraction success rate against tolerant extraction configuration.}
\label{fig:tolerance-success}
    \label{fig:tolerance-success}
\end{figure*}

\textbf{Zhao et al. Tolerant Extraction.}
Zhao et al. propose five robustness improvements to the basic extraction procedure. We implement two: NSD degree bucketization ($B>1$), which groups neighbor degrees into fixed-width bins so that small degree perturbations do not shift a node's NSD label, and approximate subgraph matching ($L>0$), which accepts a candidate subgraph if it differs from the regenerated watermark by at most $L$ edges. We adopt the parameter values from Zhao et al.'s own experimental evaluation: $B=10$ and $L=8$. All other settings match the strict configuration. The remaining three improvements (approximate NSD label matching, redundant embedding across disjoint subgraphs, and hierarchical group watermarking) substantially increase per-trial extraction cost and were computationally prohibitive given our repeated-extraction evaluation protocol; our goal is to characterize how cluster-aware attacks behave under a more permissive extraction regime rather than to evaluate maximal robustness.

Figure~\ref{fig:tolerance-success} reports extraction success rate under the tolerant configuration across the same attack strategies, datasets, and flip budgets evaluated in Section~\ref{results}. Structural distortion is omitted because tolerance affects only the extraction decision, so edit distance, dK-2 deviation, and clustering coefficient change are identical to Figure~\ref{fig:zhao_structural_distortion}. Tolerant extraction preserves the core finding from the strict configuration: cluster-aware attacks reduce extraction success with fewer edge modifications than the random baseline across all three datasets and both strategies. The tolerance regime extends the flip budget required to reach 0\% extraction but leaves the relative ordering of attack methods unchanged. Greedy Modularity and Leiden remain the most consistent cluster-aware methods, and the random baseline continues to require the largest perturbation budget to defeat the watermark on every dataset, confirming that the vulnerability to structure-aware adversaries is not an artifact of strict extraction.

\section{Runtime Performance}~\label{runtime-performance}
To assess computation, we measure clustering runtime (time to partition the graph into communities) and attack runtime (time to execute edge modifications). For cluster-aware attacks, total runtime is the sum of both components; the random baseline incurs only attack runtime. We report averages across 10 trials. Table~\ref{tab:clustering_performance} reports clustering performance. Greedy Modularity is the clear outlier, requiring hundreds of seconds on the larger graphs while Label Propagation, Leiden, and Infomap complete in under 2 seconds across all datasets. 

\begin{table}[t]
\centering
\scriptsize
\begin{tabular}{lcccccc}
\toprule
& \multicolumn{2}{c}{\textbf{Facebook}} & \multicolumn{2}{c}{\textbf{CAIDA}} & \multicolumn{2}{c}{\textbf{ArXiv}} \\
\cmidrule(lr){2-3} \cmidrule(lr){4-5} \cmidrule(lr){6-7}
\textbf{Clustering Algorithm} & \textbf{\# Clusters} & \textbf{Time (s)} & \textbf{\# Clusters} & \textbf{Time (s)} & \textbf{\# Clusters} & \textbf{Time (s)} \\
\midrule
Greedy Modularity & 16 & 19.855 & 51 & 222.627 & 452 & 461.795 \\
Label Propagation & 44 & 0.212 & 2,224 & 1.572 & 1,017 & 1.814 \\
Leiden & 17 & 0.393 & 42 & 0.579 & 324 & 1.389 \\
Infomap & 75 & 0.356 & 1,148 & 0.712 & 951 & 1.314 \\
\bottomrule
\end{tabular}
\caption{Clustering algorithm performance. We report the number of detected communities and the average clustering runtime (in seconds).}
\label{tab:clustering_performance}
\end{table}

\begin{table}[t]
\centering
\scriptsize
\begin{tabular}{lccc}
\toprule
\textbf{Attack Strategy} & \textbf{Facebook} & \textbf{CAIDA} & \textbf{ArXiv} \\
\midrule
Random Baseline & 1.45$\pm$0.01 & 45.65$\pm$0.18 & 23.04$\pm$0.11 \\
Intra-Add/Inter-Remove & 2.43$\pm$0.07 & 72.79$\pm$1.82 & 40.05$\pm$4.69 \\
Intra-Remove/Inter-Add & 3.41$\pm$0.10 & 139.87$\pm$23.29 & 66.31$\pm$4.19 \\
\bottomrule
\end{tabular}
\caption{Average time (in seconds) required to execute each attack strategy.}
\label{tab:attack_runtime}
\end{table}

Table~\ref{tab:attack_runtime} reports attack execution times excluding clustering overhead. Cluster-aware attacks incur modest additional cost from checking community membership per modification, with total runtime ranging from under 5 seconds on \textit{Facebook} to roughly 600 seconds on \textit{ArXiv} with Greedy Modularity. These results confirm cluster-aware attacks remain practical for offline adversaries.

\end{document}